\documentclass[a4paper,fleqn,usenatbib]{mnras}

\usepackage{mathptmx}
\usepackage{txfonts}

\usepackage[T1]{fontenc}
\usepackage{ae,aecompl}

\usepackage{graphicx}	% Including figure files
\usepackage{amsmath}	% Advanced maths commands
\usepackage{amssymb}	% Extra maths symbols

\title[IGR J19294$+$1816: a new BeXB]{IGR J19294$+$1816: a new Be-X ray binary revealed
through infrared spectroscopy}

\author[J. J. Rodes-Roca et al.]{
J. J. Rodes-Roca,$^{1,2}$\thanks{E-mail: jjrodes@ua.es}
G. Bernabeu,$^{1,2}$
A. Magazz\`u,$^{3}$ 
J. M. Torrej\'on$^{1,2}$\newauthor
and E. Solano$^{4,5}$
\\
$^{1}$Department of Physics, Systems Engineering and Signal Theory, University of Alicante,
              03080 Alicante, Spain\\
$^{2}$University Institute of Physics Applied to Sciences and Technologies,
              University of Alicante, 03080 Alicante, Spain\\
$^{3}$Telescopio Nazionale Galileo, Rambla Jos\'e Ana Fern\'andez P\'erez, 38712 Bre\~na Baja, Spain\\
$^{4}$Departamento de Astrof\'{\i}sica, CAB (CSIC-INTA), ESA-ESAC Camino Bajo del Castillo s/n, 28692 Villanueva de la Ca\~nada, Madrid, Spain\\
$^{5}$Spanish Virtual Observatory
}

\date{Accepted XXX. Received YYY; in original form ZZZ}

\pubyear{2017}

\begin{document}
\label{firstpage}
\pagerange{\pageref{firstpage}--\pageref{lastpage}}
\maketitle

% Abstract of the paper
\begin{abstract}
The aim of this work is to characterize the counterpart to the \emph{INTEGRAL}
High Mass X-ray Binary candidate IGR J19294+1816 so as to establish its true nature. 
We obtained $H$ band spectra of
the selected counterpart acquired with the \emph{NICS}
instrument mounted on the \emph{Telescopio Nazionale Galileo} (TNG) 3.5-m
telescope which represents the first infrared spectrum ever taken of this source. We complement the spectral analysis with infrared photometry from
UKIDSS, 2MASS, WISE and NEOWISE databases. We classify the mass donor as a
Be star. Subsequently, we compute its distance by properly taking into account the contamination produced by the circumstellar envelope. The findings indicate that IGR J19294+1816 is a transient source with a B1Ve donor at a distance of $d=11\pm1$ kpc, and luminosities of the order of
$10^{36-37}$ erg s$^{-1}$, displaying the typical behaviour of a Be
X-ray binary.
\end{abstract}

% Select between one and six entries from the list of approved keywords.
% Don't make up new ones.

%%%%%%%%%%%%%%%%%%%%%%%%%%%%%%%%%%%%%%%%%%%%%%%%%%

%%%%%%%%%%%%%%%%% BODY OF PAPER %%%%%%%%%%%%%%%%%%

\section{Introduction}

Current X-ray observatories such as the European Space Agency's (ESA)
\emph{INTErnational Gamma-Ray Astrophysics Laboratory} (INTEGRAL;
\citeauthor{2003A&A...411L...1W} 2003) or \emph{XMM-Newton}
(\citeauthor{2009A&A...493..339W} 2009)
have discovered new kinds of high-energy emitters. To
characterize the nature of these systems it is mandatory
to carry out a multiwavelength study and establish a clear identification of the optical/infrared counterpart.

High-mass X-ray binaries (HMXBs) are X-ray sources fed by accretion
of material from a donor OB star onto a compact object (a black
hole or a neutron star). They are primary astrophysical laboratories
where fundamental properties can be tested, such as the following: the masses of neutron stars and the equation state of the nuclear matter; the structure of the stellar wind in massive stars; or, the evolutionary paths of binary systems.\citep{2017SSRv..tmp...13M}. %silvia issi paper
Until recently, three main kinds of HMXBs were known:
\begin{enumerate}
    \item The largest group has a Be star as a donor. These systems
are called BeX systems \citep{2011Ap&SS.332....1R}. A Be star is a main sequence or giant star which
harbours a circumstellar equatorial disk. These systems are transient
(in X-rays) with a duration of some weeks and show luminosities
of $L_X\sim 10^{36}-10^{37}$ erg\,s$^{-1}$. Approximately 70\% of all known X-ray pulsars belong to this class. 
    \item An increasing number of sources have a supergiant OB donor (SGXBs or
SFXTs). SGXBs are persistent sources where the compact object accretes material
from the powerful stellar wind, producing luminosities
of the order of $L_X\sim 10^{36}$ erg\,s$^{-1}$. SFXTs are transient
sources showing luminosities of $L_X\sim 10^{33-34}$ erg\,s$^{-1}$
during quiescence that can increase by up to four orders of magnitude
in very short time scales.
    \item A very small group where the transfer of mass is directly
from the surface of the donor star (Roche Lobe overflow). This 
evolutionary phase is very short and is consequently
very scarce. For instance, in the Milky Way only the Cen X-3 is presently observable.
\end{enumerate}
The population synthesis models that were previously available seemed to reproduce the distribution well. For example, the brevity of the supergiant phase was reflected in the small number of SGXBs detected. However, with the discovery of an increasing
number of obscured sources by high energy satellites like ESA's
\emph{INTEGRAL} gamma ray telescope, the population synthesis models started to change. Given the \emph{INTEGRAL}'s sensitivity above
20 keV and its observing strategy, which produce very long exposure
times specifically towards the galactic centre, new sources were discovered, that had been missed in the past due to the very high absorption or the very short transient nature. The majority of these newly discovered sources had supergiant donors. The persistent sources dramatically increased the number of SGXBs. As an example, we 
identified the \emph{XMM-Newton} source
2XMM J191043.4+091629 with a distant SGXB \citep{2013A&A...555A.115R}
which helps to independently trace the galactic structure. It also contains the slowest pulsar found to date \citep{2017MNRAS.469.3056S}. Furthermore, the transient systems revealed an entirely new class of objects, the Supergiant Fast X-ray Transients
(SFXTs). Finally, we have discovered obscured BeX systems
(main sequence donors) close to earth. All these discoveries can
challenge the population synthesis models. Therefore, to
characterize as many counterparts as possible is very important.

Often these new sources turn out to be highly absorbed. The
spectral classification using the blue band would require very long
exposure times on large telescopes or it is not feasible. 
However, they can be perfectly observed with the
IR instrumentation on a 4-m class telescope.
The spectral classification of hot stars can be achieved with
$H$-band spectra \citep{1998AJ....116.1915H}. Together with
available infrared photometry and the X-ray behaviour, the nature of
the system can be established unambiguously.

IGR J19294$+$1816 was discovered by the \emph{IBIS/ISGRI} imager
on-board \emph{INTEGRAL} during an observation of the field
around GRS 1915+105 on March 27, 2009 
\citep{2009ATel.1997....1T}. %turler et al. 2009Atel
Using \emph{Swift/XRT} archival observations,
\citet{2009ATel.1998....1R} %rodriguez et al. 2009Atel
gave a refined X-ray position identified with the
source named Swift J1929.8$+$1818 ($\alpha$ = 19$^h$ 29$^m$ 56$^s$
and $\delta =$ +18$^\circ$ 18$^\prime$ 38$^{\prime\prime}$ with an uncertainty of 3.5 arcseconds
at 90\% confidence level). The single star falling in the \emph{XRT} error circle was identified as the IR counterpart to the X-ray source, and associated with 2MASS J19295591$+$1818382. However, no optical or radio counterpart was found in any other catalogues they searched for. The \emph{XRT} spectrum
is well described by an absorbed power law or by a single
absorbed blackbody. The timing analysis revealed the presence
of a pulse around 12.43 s which was later confirmed with \emph{RXTE}
\citep{2009ATel.2002....1S}. %strohmayer et al. 2009Atel
The X-ray source displays short outbursts (several thousand seconds)
which suggested initially a SFXT nature \citep{2009A&A...508..889R}.
\citet{2011A&A...531A..65B}, %bozzo et al. 2011
from \emph{INTEGRAL} and \emph{Swift} observations, concluded
that the X-ray source showed a behaviour reminiscent of a Be X-ray binary system.

In this work, we present the first ever infrared spectrum of IGR
J19294$+$1816\footnote{http://irfu.cea.fr/Sap/IGR-Sources} supported
by the analysis of archival IR photometry in order to ascertain the
true nature of the source. In
Section~\ref{data}, we describe the
IR observations. In Section~\ref{analyse}, we present our results and,
finally, we present our conclusions in Section~\ref{conclusion}.

%with small
%position error (provided
%by X-ray telescopes like \emph{XMM-Newton} or \emph{Swift}).
%Using Virtual Observatory (VO) tools such as
%\textsc{CDS ALADIN} \citep{2000A&AS..143...33B} and \textsc{TOPCAT}
%\citep{2005ASPC..347...29T}, we have confirmed the potential candidate
%counterpart.

\section{Observations and data reduction}
\label{data}

\subsection{TNG observations}

Near-IR spectroscopy was obtained
during the night of September 1st, 2014, using the Near Infrared
Camera and Spectrograph (NICS) mounted at the 3.5-m Telescopio
Nazionale Galileo (TNG) 
telescope (La Palma island). Medium-resolution spectra were taken with the
H grism under good seeing conditions and reasonable signal-to-noise
(S/N) ratio of 100. 

To remove the sky background, the target and the standard stars
were observed according to a nodding ABBA sequence along the slit, using an automatic
script available at the telescope. Consequently, each
observation consisted of four images with the source spectrum displaced at
different positions on the detector. The separation between the A and B
positions was $15^{\prime\prime}$.

As the first step of the reduction process, possible cross-talking effects
were removed using a Fortran program
available from the TNG web pages.
Second, background subtraction was made
by taking (A$-$B) and (B$-$A) image differences, obtaining four positive
aperture images. The AB and BA sequences were so close in time
that sky background variation between them was negligible.
This method, together with the use of the 1.0$^{\prime\prime}$ slit, also
minimizes any possible nebular contamination.

For each differential image, a spectrum was extracted using the \emph{apall}
task in the \emph{IRAF}\footnote{Image Reduction and Analysis Facility is written and
supported by the National Optical Astronomy Observatories
which is operated by the Association of Universities for
Research in Astronomy, Inc. under cooperative agreement
with the National Science Foundation} 
environment. During the extraction, we removed the residual background
and traced the apertures along the dispersion. The four extracted spectra of
each group were then combined together with a median algorithm to produce
a single spectrum, thereby eliminating cosmic ray spikes.

To remove the telluric features, a number of A0V stars were observed
throughout the night at similar airmasses as the targets. These
spectra were fitted with a theoretical spectrum corrected for rotation
and radial velocity. The standard spectra were divided by the model
isolating the telluric spectrum which was used subsequently to correct
the target spectra. We carefully checked that division by the
telluric spectrum does not introduce any spurious features. At our
resolution (3.5 \AA/pix), this method works very well for the $H$ band. However, it could somewhat increase the emission lines seen in the $K$ band, although we estimate this effect to be smaller than 20\% in
Br$\gamma$. In this paper we are concerned with the characterisation
of the donor. As will be shown, it turns out to be a Be
star. In the IR essentially, we are seeing the circumstellar envelope
emission and, consequently, the line ratios cannot be used for
spectral classification. Therefore we have not corrected further for this effect\footnote{The corrected spectra will be used to diagnose specifically the physical properties of the circumstellar envelopes. However, this is beyond the scope of this paper}.
The final spectra are shown in Fig.~\ref{HKband}.

\section{Data analysis}
\label{analyse}

\subsection{Near-IR spectra and classification of the counterpart}
\label{nir}

The spectral analysis was carried out using the
\emph{Starlink}\footnote{http://starlink.jach.hawaii.edu/starlink} software
and the \emph{IRAF} package\footnote{http://iraf.noao.edu}.
To identify the emission/absorption lines and
spectral classification, we used the following atlases:
\citet{1997AJ....113.1855B}, \citet{1998ApJ...508..397M} and
\citet{1998AJ....116.1915H} for
the H band; \citet{1996ApJS..107..281H} and \citet{2005ApJS..161..154H} for
the K band.

\begin{figure*}
  \centering
  \includegraphics[angle=0,width=\columnwidth]{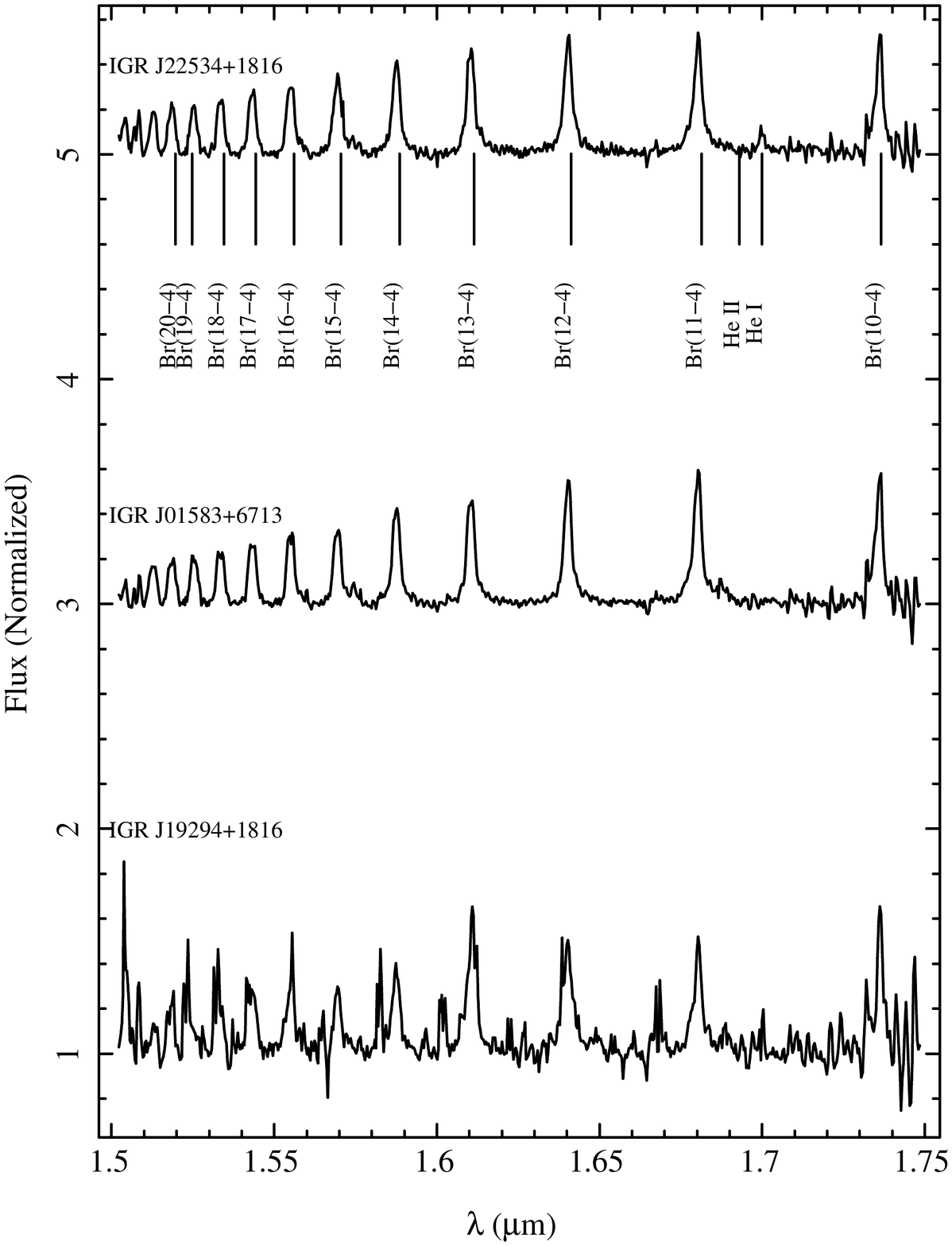}
  \includegraphics[angle=0,width=\columnwidth]{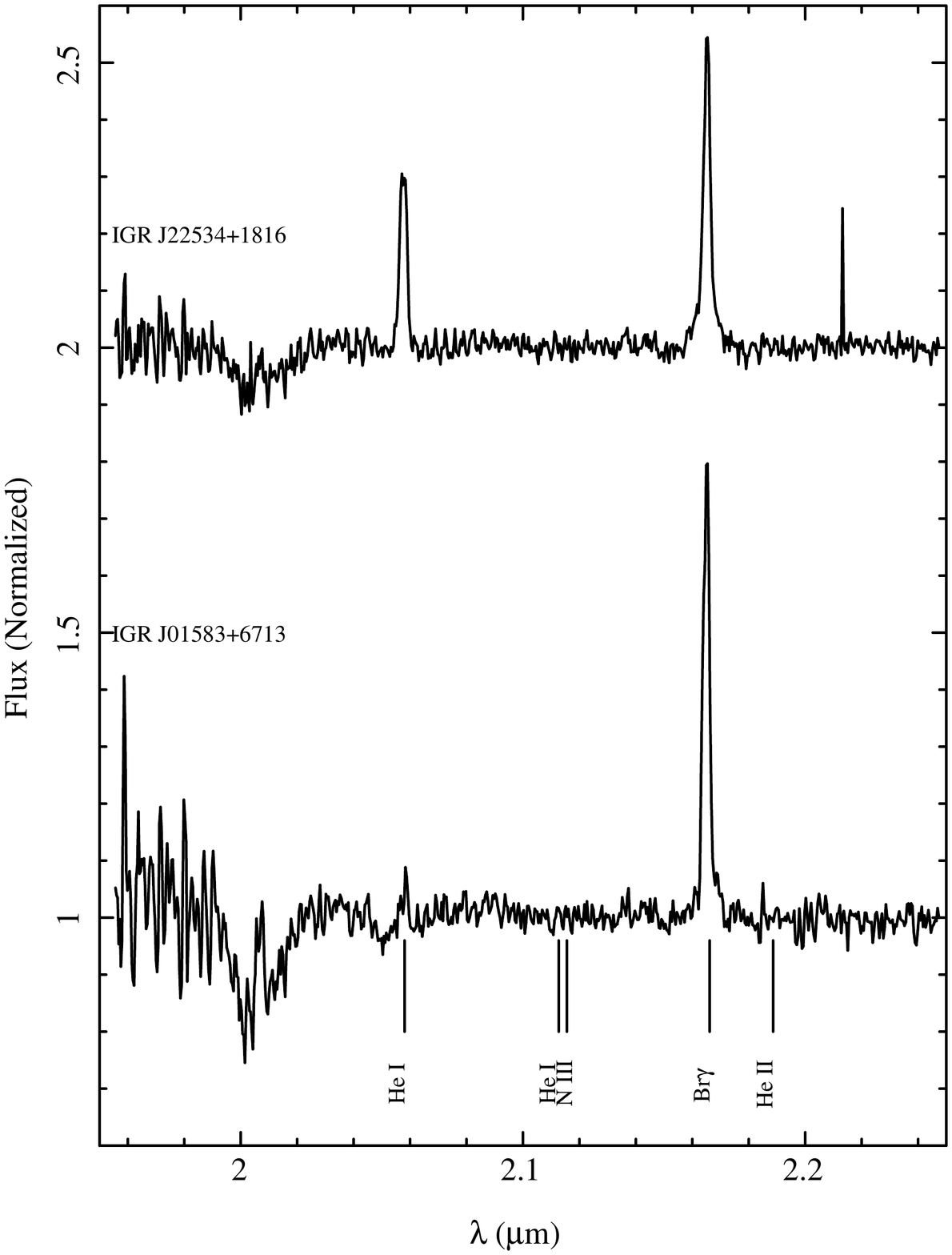}
  \caption{NIR spectra of the counterparts to the X-ray
binaries. No other emission or absorption lines were detected apart from
Brackett series of H \textsc{i} and He \textsc{i}
2.058 $\mu$m emission lines. Note  the absence of any feature
at the position of the He \textsc{ii} 2.1885 $\mu$m line.
  }
  \label{HKband}
\end{figure*}

Figure~\ref{HKband} shows our $H$ band \textit{TNG} spectrum. To date, this is the only infrared spectrum obtained for this source. The strong absorption (see section
\ref{IR}) $A_V^{tot} = 15.7 = 3.1\, E(B-V)$
implies a colour excess of $E(B-V) = 5.1$ and makes it difficult to
obtain an optical blue spectrum with a 4-m class telescope.
The NIR $H$-band spectrum exhibits the presence of the Brackett H \textsc{i} series
from Br(19-4) line at 1.5235 $\mu$m to the Br(10-4) transition at
1.7362 $\mu$m. This points clearly  towards
an early B type star, since these lines disappear for O type
stars. All these lines are in emission. The 1.700 $\mu$m He \textsc{i} line is seen also in
emission. This is typical of Be stars in which the circumstellar disk
emission dominates the spectrum at NIR wavelengths. Some Brackett
series lines show a double peak structure.
According to a few analyses of infrared emission line profiles in Be stars,
they have the same characteristics as those in the visible region
\citep{1982JBAA...92..290U}. Most Be stars show double-peak structure
in their Balmer emission lines \citep{1992ApJS...81..335S}. In general,
this double-peak structure is to be expected for a rotating emitting disk.
If the star is viewed at an intermediate angle of inclination of
the line of sight on the rotation axis, the emission lines exhibit
a weak central reversal typical of a Be spectrum.
Features at 1.5965 $\mu$m [Fe \textsc{ii}], at 1.6027 $\mu$m C \textsc{i}
(possibly blended with 1.6002 $\mu$m [Fe \textsc{ii}] and/or 1.6009 $\mu$m C \textsc{i})
and at 1.6685 $\mu$m Si \textsc{i} seem to be present in our
spectrum. Another line could match the position of He \textsc{ii} at
1.697 $\mu$m. He \textsc{ii}
is only detected in O stars, and preferentially in the supergiants
\citep{1998AJ....116.1915H}. %hanson et al. 1998. 
However, He \textsc{ii} is very weak preventing differentiation from the
continuum noise level. Furthermore, $H$-band spectra of Be stars show all Brackett
series in emission \citet{2001A&A...371..643S}. %steele and clark
At any rate, the $H$ band spectrum seems to be
characteristic of a Be star. 

As a comparison, we also show for the first time NIR spectra of two
previously classified Be X-ray binaries based on optical spectra. For IGR J01583$+$6713, \citet{2005ATel..681....1H} %hapern & Tyagi 2005Atel
obtained a low-resolution optical spectrum which
revealed strong H$\alpha$ and weak H$\beta$
emission lines, of equivalent width (EW) 70~\AA\ and
6~\AA\ respectively, as well as diffuse interstellar bands.
These features suggested a Be companion. 
Subsequent optical spectroscopy \citep{2006A&A...455...11M, 2008MNRAS.386.2253K} %masseti et al. 2006
confirmed its Be-X nature. For IGR J22534$+$6243,
\citet{2012ATel.4248....1M} %masetti et al. 2012Atel
obtained a spectrum of a highly reddened, intrinsically blue
continuum star with superimposed H$\alpha$, H$\beta$ and He \textsc{i}
emission lines pointing strongly to a Be companion. 
\citet{2013MNRAS.433.2028E} %esposito et al. 2013
confirmed the Be-X nature using a blue band spectrum of the IR
counterpart obtained with the 2.5-m Nordic Optical Telescope (NOT, La Palma).
Although the $S/N$ is poorer for IGR J19294$+$1816 (because it is 2
magnitudes fainter in $H$) it is clearly consistent with a Be donor. 

Table~\ref{tab:ew} summarises the line identifications and the estimated
equivalent widths for all the observed targets.

\begin{table*}
\caption{$H-$ and $K-$band line identifications for each source which the corresponding
equivalent widths.}
\centering
\begin{tabular}{lcccc}
\hline
\hline
&  & \multicolumn{3}{c}{IGR J} \\
Emission & Wavelength & 01583+6713 & 19294+1816 & 22534+6243 \\
line & ($\mu m$) & \multicolumn{3}{c}{EW (\AA)} \\
\hline
Br 20 & 1.5196 & -5.5$\pm$0.8 & -5.5$\pm$0.8 & -5.9$\pm$0.8 \\
Br 19 & 1.5265 & -5.7$\pm$0.8 & -9.3$\pm$1.3 & -5.5$\pm$0.8 \\
Br 18 & 1.5346 & -7.0$\pm$1.0 & -9.5$\pm$1.4 & -7.0$\pm$1.0 \\
Br 17 & 1.5443 & -8.2$\pm$1.2 & -10.6$\pm$1.5 & -8.5$\pm$1.2 \\
Br 16 & 1.5561 & -9.9$\pm$1.4 & -10.8$\pm$1.5 & -10.3$\pm$1.5 \\
Br 15 & 1.5705 & -10.5$\pm$1.5 & -7.5$\pm$1.1 & -11.4$\pm$1.6 \\
Br 14 & 1.5885 & -14.8$\pm$2.1 & -11.6$\pm$1.7 & -14.7$\pm$2.1 \\
Fe \textsc{ii} & 1.5965 & --- & -1.4$\pm$0.2 & --- \\
Fe \textsc{ii} & 1.6019 & --- & blended? & --- \\
C \textsc{i} & 1.6027 & --- & -5.1$\pm$0.7 & --- \\
Br 13 & 1.6114 & -14.9$\pm$2.1 & -20$\pm$3 & -16.7$\pm$2.4 \\
Br 12 & 1.6412 & -19$\pm$3 & -21$\pm$3 & -19$\pm$3 \\
Si \textsc{i} & 1.6685 & --- & -4.7$\pm$0.7 & --- \\
Br 11 & 1.6811 & -22$\pm$3 & -17.4$\pm$2.4 & -21$\pm$3 \\
He \textsc{i} & 1.7007 & --- & -1.4$\pm$0.2 & -2.3$\pm$0.3 \\
Br 10 & 1.7362 & -19$\pm$3 & -14.9$\pm$2.1 & -16.2$\pm$2.3 \\
He \textsc{i} & 2.0580 & -1.3$\pm$0.2 & --- & -9.9$\pm$1.4 \\
He \textsc{i} & 2.1126 & < -0.2 & --- & < -0.3 \\
N \textsc{iii} & 2.1155 & < -0.2 & --- & < -0.1 \\
Br $\gamma$ & 2.1661 & -24$\pm$3 & --- & -20$\pm$3 \\
He \textsc{ii} & 2.1885 & < -0.2 & --- & < -0.4 \\
\hline
\hline
\end{tabular}
\label{tab:ew}
\end{table*}

\begin{figure}
  \centering
  \includegraphics[width=\columnwidth]{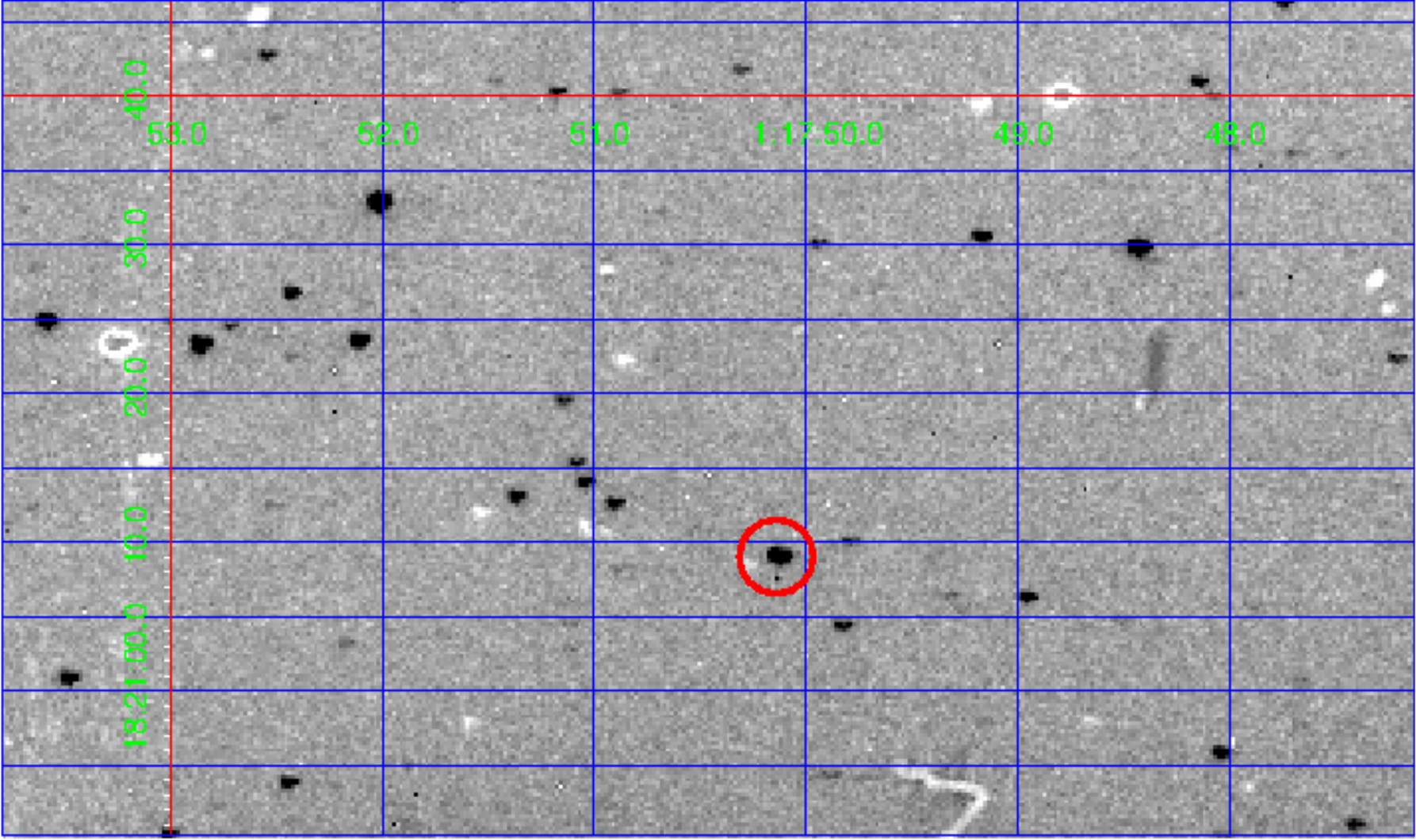}
  \caption{TNG $K$-band image of the region with the counterpart marked
(red circle). The white dots are ghosts due to imperfect subtraction
of the images during dithering. However, the image was used only for
object identification and not for photometric analysis. North is up,
East is left.
  }
  \label{fig:kimage}
\end{figure}

In order to rule out any contamination by an unresolved companion,
we compared the same field of view in the $K$-band
of 2MASS and UKIDSS as this last survey provides a higher angular
resolution. This is a necessary test for faint sources (typically
$K > 10 \, \,\text{mag}$) as we demonstrated in
\citet{2013A&A...555A.115R}. The source is not blended, as
it shown in Fig.~\ref{fig:kimage}.

In summary, the NIR spectrum of IGR J19294+1816 points strongly
towards a Be nature of the
donor. This conclusion is further strengthened
by the photometry presented in the next section. 

%Moreover,
%the direct application of the usual photometric techniques
%to calculate the distance to Be X-ray binaries might lead to incorrect results
%\citep{2012A&A...539A.114R}. %riquelme et al. 2012

\subsection{Photometry}
\label{IR}

To carry out the photometric analysis we used
the Simbad database \citep{2000A&AS..143....9W}, % Wenger et al. 2000
several optical databases, such as \emph{Gaia}
\citep{2016A&A...595A...2G,2016A&A...595A...1G,2017A&A...599A..32V,
2017A&A...600A..51E}, \emph{IPHAS} \citep{2005MNRAS.362..753D},
and \emph{PAN-STARRS} \citep{2016arXiv161205560C,2016arXiv161205240M,
2016arXiv161205245W,2016arXiv161205244M,2016arXiv161205242M,2016arXiv161205243F}
and several IR databases
such as the UK Infrared Telescope (UKIRT) Infrared Deep Sky Survey (UKIDSS,
\citet{2007MNRAS.379.1599L}), % Lawrence et al. 2007
the Two Micron All Sky Survey (2MASS, \citet{2006AJ....131.1163S}), the
GLIMPSE database, the
Wide-field Infrared Survey Explorer (WISE) all sky survey \citep{2010AJ....140.1868W},
and the Near-Earth Object WISE (NEOWISE, \citet{2014ApJ...792...30M})
obtaining the photometry given in Table~\ref{photometry}.

\begin{table}
\caption{Photometry of candidate counterparts.}
\centering
\begin{tabular}{lccc}
\hline
\hline
Photometry & \multicolumn{3}{c}{IGR J} \\
(mag) & 01583+6713 & 19294+1816 & 22534+6243 \\
\hline
$B$ & 15.65$\pm$0.08 & --- & 17.38$\pm$0.11 \\
$V$ & 14.43$\pm$0.03 & --- & 15.61$\pm$0.01 \\
$B-V$ & 1.22$\pm$0.11 & --- & 1.77$\pm$0.12 \\
& \multicolumn{3}{c}{2MASS} \\
$J$ & 11.48$\pm$0.03 & 14.56$\pm$0.03 & 11.644$\pm$0.024 \\
$H$ & 11.03$\pm$0.03 & 12.99$\pm$0.03 & 10.961$\pm$0.023 \\
$K$ & 10.601$\pm$0.021 & 12.115$\pm$0.023 & 10.46$\pm$0.03 \\
& \multicolumn{3}{c}{UKIDSS} \\
$J$ & --- & 14.3019$\pm$0.0020 & --- \\
$H$ & --- & 12.8710$\pm$0.0009 & --- \\
$K$ & --- & 11.8792$\pm$0.0011 & --- \\
& \multicolumn{3}{c}{WISE/NEOWISE} \\
$W1$ & 10.099$\pm$0.023 & 10.715$\pm$0.024 & 9.87$\pm$0.04 \\
$W2$ & 9.715$\pm$0.020 & 10.283$\pm$0.022 & 9.45$\pm$0.03 \\
$W3$ & 8.85$\pm$0.03 & 10.27$\pm$0.15 & 8.57$\pm$0.05 \\
$W4$ & 8.4$\pm$0.3 & <8.74 & 7.73$\pm$0.19 \\
& \multicolumn{3}{c}{GLIMPSE} \\
3.6 $\mu m$ & --- & 10.95$\pm$0.05 & --- \\
4.5 $\mu m$ & --- & 10.66$\pm$0.05 & --- \\
5.8 $\mu m$ & --- & 10.39$\pm$0.07 & --- \\
8.0 $\mu m$ & --- & 10.20$\pm$0.04 & --- \\
& \multicolumn{3}{c}{PAN-STARRS} \\
$g$ (0.477 $\mu m$) & --- & 27.06 & --- \\
$r$ (0.613 $\mu m$) & --- & 22.29$\pm$0.16 & --- \\
$i$ (0.748 $\mu m$) & --- & 20.15$\pm$0.05 & --- \\
$z$ (0.865 $\mu m$) & --- & 18.335$\pm$0.014 & --- \\
$y$ (0.960 $\mu m$) & --- & 17.223$\pm$0.014 & --- \\
& \multicolumn{3}{c}{Gaia} \\
$G$ (0.673 $\mu m$) & --- & 20.71 & --- \\
& \multicolumn{3}{c}{IPHAS} \\
$i$ (0.766 $\mu m$) & --- & 19.51$\pm$0.08 & --- \\
\hline
\hline
\end{tabular}
\label{photometry}
\end{table}

\begin{figure}
  \centering
  \includegraphics[angle=-90,width=\columnwidth]{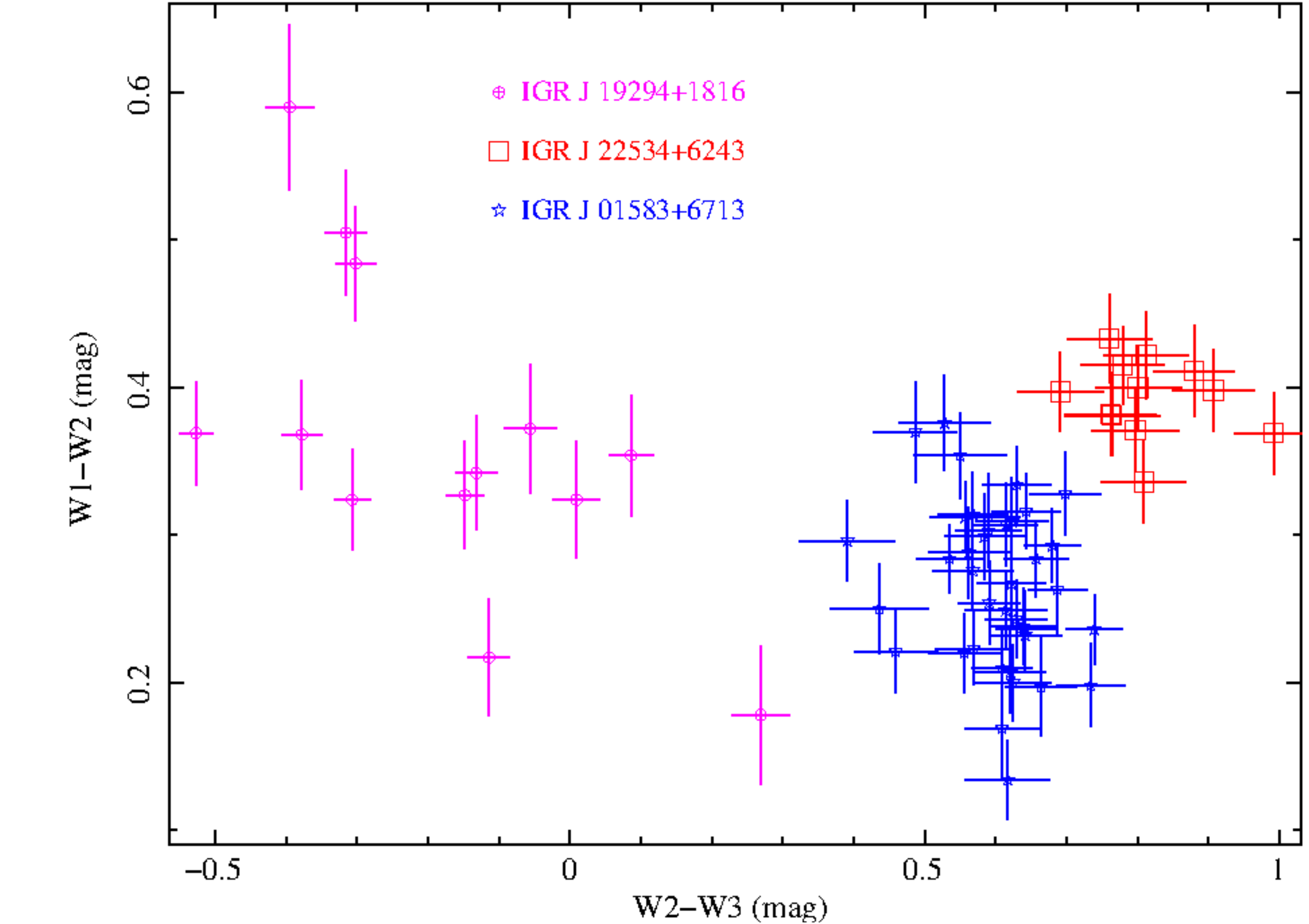}
  \caption{WISE All-Sky Single Exposure photometry colour-colour diagram.
  }
  \label{fig:wise}
\end{figure}

Classical Be stars tend to occupy a reduced region in the colour-colour (CC)
diagram [W1$-$W2] vs [W2$-$W3], see figure~5
and figure~7 central panel in \citet{2014ApJ...791..131K}. % koenig - wise
They are located in the range [-0.1--0.6] vs [-0.1--1.5] in this diagram.
We obtained the single exposure data from the WISE All-Sky Single
Exposure (L1b) database.
Then, we selected cc\_flag equal to 0 or h and non-null data for W3
and did the previous CC diagram on the \textit{IRSA} (Infrared Science Archive) web
application (Fig.~\ref{fig:wise}).
For the Be/X-ray binaries IGR J01583+6713 and IGR J22534+6243
all the values are
found in the classical Be zone\footnote{although can be confused with blue
transition disk objects}. No significant variability was found
within the uncertainties. For IGR J19294+1816 the values of the W3 filter had
a spurious detection and only an upper limit on magnitude could be estimated.
Therefore, error bars in $W2-W3$ are the uncertainties measured in $W2$.
Nevertheless, the points in the CC diagram
were also consistent with a classical Be star.

\begin{figure}
  \centering
  \includegraphics[angle=-90,width=\columnwidth]{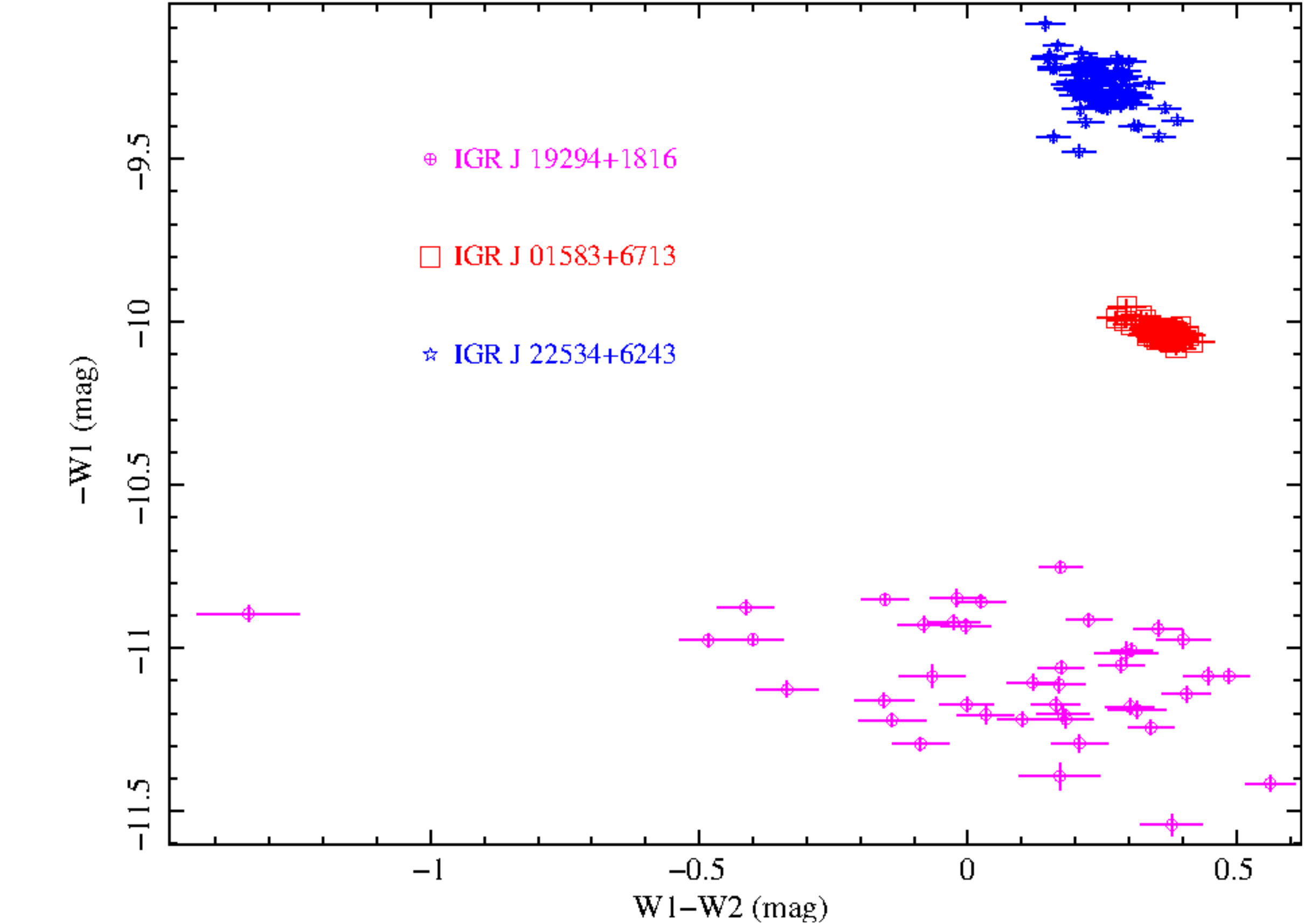}
  \caption{NeoWISE multiepoch photometry colour-magnitude diagram.
  }
  \label{fig:neowise}
\end{figure}

\begin{figure}
  \centering
  \includegraphics[angle=-90,width=\columnwidth]{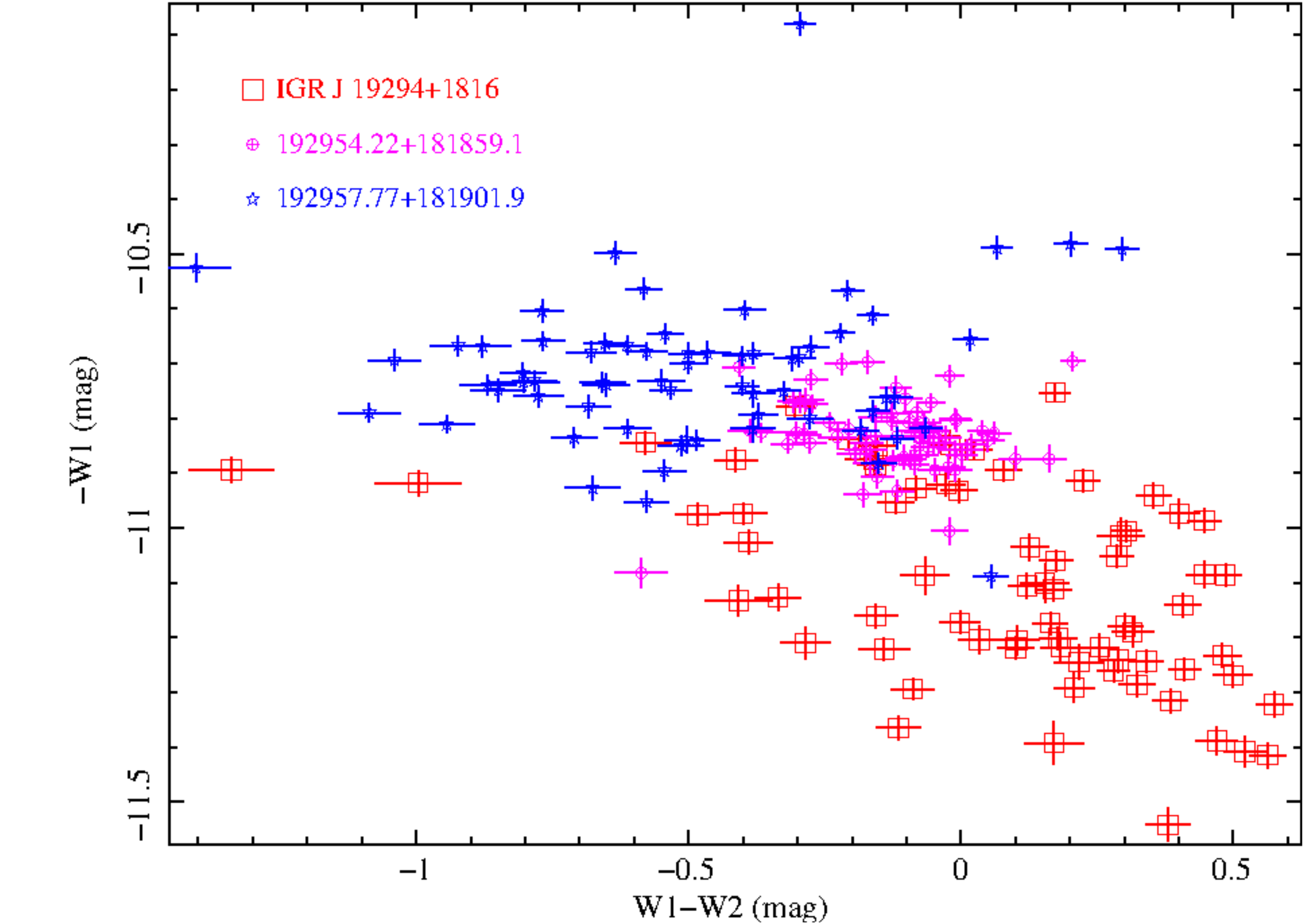}
  \caption{NeoWISE multiepoch photometry colour-magnitude diagram of IGR J19294+1816
and two nearby sources.
  }
  \label{fig:neowise2}
\end{figure}

In addition, the \textit{NeoWISE} database has single exposure data for these objects
allowing us to plot the colour-magnitude (CM) diagram [$-$W1] vs [W1$-$W2]
(Fig.~\ref{fig:neowise}). IGR J19294+1816 seems to be quite
variable. However, we built the same CM diagram for two nearby sources and also
showed the same behaviour (Fig.~\ref{fig:neowise2}).
The standard deviation of the variability in the WISE colour is
$\overline{W_1-W_2}_{\mathrm{(IGR)}} = 0.1\pm0.3$,
$\overline{W_1-W_2}_{\mathrm{(Source1)}} = -0.12\pm0.14$,
$\overline{W_1-W_2}_{\mathrm{(Source2)}} = -0.5\pm0.3$, respectively.

%At this point, we are not sure if this is a real
%variability or it is due to a transient effect.


As seen in Section \ref{analyse} all IR lines are in emission. The circumstellar envelope emission contaminates the underlying
photospheric spectrum of the Be star and produces an
overluminosity (for non Be-shell) with respect to B stars of the same
spectral type and luminosity class. This leads to a systematic under
estimation of the distance and, hence, the X-ray luminosity.
We will use the recipes given in \citet{2012A&A...539A.114R} to
correct for these effects. The distance is computed from the usual
equation 

\begin{equation}
V-M_{V}-A^\text{tot}_{V}=5\log d -5
\label{eq:distance}
\end{equation}
where $V$ is the observed magnitude, $M_{V}$ is the absolute magnitude
for the corresponding spectral type and luminosity class and $A^\text{tot}_{V}$
has two terms: the interstellar absorption towards the
source $A^\text{is}=3.1\, E^\text{is}(B-V)$ and the circumstellar
\emph{emission}:

\begin{equation}
A^\text{tot}_\text{V}=\left\{
\begin{array}{ll}
3.1\, E^\text{is} (B-V) - 0.6\ \left(\frac{W^{\text{env}}(\text{H}_{\alpha})}{-30 \mathring{A}}\right) & \mbox{if } -W^{\text{env}}(\text{H}_{\alpha}) < 15 \mathring{A} \\
3.1\, E^\text{is} (B-V) - 0.3 & \mbox{if } -W^{\text{env}}(\text{H}_{\alpha}) \geq 15 \mathring{A} 
\end{array}
\right .
\label{eq:excess}
\end{equation}

The fact that all the significant lines are in emission implies a well developed
envelope which is responsible for the strong IR excess.
Both, IGR J01583+6713 \citep{2005ATel..681....1H} %hapern & Tyagi 2005Atel
and IGR J22534+6243 \citep{2013MNRAS.433.2028E} %esposito et al. 2013
have values of measured $-W
(\text{H}_{\alpha})$ larger than 15 \AA\ (Table~\ref{tab:ew}) and, therefore, they are in the
saturated regime. On the other hand, IGR J19294+1816 presents also all
the significant $H$ band lines in emission, with comparable equivalent
widths, demonstrating that the
envelope is well developed. We will then assume that this source is
also in the saturated regime. In these conditions: 

\begin{equation}
A_V^\text{tot} = 3.1\, E^\text{is}(B-V) - 0.3
\label{eq:excess2}
\end{equation}

IGR J19294+1816 has no $UBV$ optical photometry due to the very
high extinction. We have derived the excess using the available
IR photometry. Be donors in Be-X-ray binaries occupy a rather narrow
spectral class interval \citep{2012A&A...539A.114R}.  We have estimated the distance assuming a B0 V or B2 V spectral
type. In \citet{2012A&A...539A.114R} it is shown that the visual to
infrared colour circumstellar excess $E(V-I)$ is already 3 to 4 times
that of $(B-V)$ for Be-X ray binaries and increases towards longer wavelengths. We estimate
that the effect on $(J-K)$ is of the order of 1 magnitude \citep{2012A&A...539A.114R}. %riquelme et al. 2012
That is to say $E^\text{cs}(J-K)\approx 1$ mag. Therefore $E^\text{is}(J-K)\approx
E^\text{tot}(J-K)-1$. Following \citet{1999PASP..111...63F} we
can compute subsequently $E^\text{is}(J-K)=0.5\, E^\text{is}(B-V)$ and
$A^\text{is}_{K}=0.36\, E^\text{is}(B-V)$. In Table~\ref{tab:extinct} we
summarize our results. As can be seen, the distances to IGR
J19294+1816 are in the range 10-12 kpc. We will assign finally a
distance $d=11\pm 1$ kpc locating the system at the far edge of the
Perseus arm. The line of sight passes through the Sagitarius arm
tangentially and the Perseus arm, explaining the high
extinction.
Using the values from Table 4 in \citet{2012A&A...539A.114R} %riquelme et al. 2012
and equation~\ref{eq:distance}, we estimate a lower limit for the visual
magnitude of $V \sim 23.4$ mag. This is corroborated by the SED
analysis in the next section.

\begin{table*}
\caption{Extinctions and distances for the IGR J19294+1816 counterpart.
Spectral type in the Morgan-Keenan system
and intrinsic colours from \citep{1982lbg6.conf.....A} and
\citep{2001ApJ...558..309D}.}
\centering
\begin{tabular}{ccccccc}
\hline
\hline
%Source IGR J & MK & $(B-V)_0$ & $E^{\text{tot}}(B-V)$ & $E^{\text{is}}(B-V)^\text{a}$ & $M_V$ & $A_V^{\text{tot}}$ (mag) & $A_V^{\text{is}}$ (mag) & $d$ (kpc) \\
%\hline
%01583+6713 & B2 IVe & $-$0.24 & 1.46 & 1.29 & $-$3.1 & 3.7 & 4.0 & 5.8 \\
%22534+6243 & B1 Ve & $-$0.3 & 5.33 & 5.16 & $-$3.2 & 5.5 & 5.8 & 4.6 \\
% & B1 IIIe & $-$0.27 & 5.23 & 5.06 & $-$4.4 & 5.5 & 5.8 & 8.0 \\
% & & & & & & & & \\ \hline
MK & $(J-K)_0$ & $E^{\text{tot}}(J-K)$ & $E^{\text{is}}(J-K)$ & $M_K$ & $A_K^{\text{is}}$ (mag) & $d$ (kpc) \\ \hline
B2 Ve & $-$0.22 & 2.665 & 1.665 & $-$3.5 & 0.30 & 11.6$^\text{b}$ \\
B0 Ve & $-$0.17 & 2.615 & 1.615 & $-$3.17 & 0.29 & 10.0$^\text{b}$ \\
\hline
\hline
\multicolumn{3}{l}{$^aE^{\text{is}}(B-V) = E^{\text{tot}}(B-V) - 0.17$} &
\multicolumn{4}{r}{$^b$Assuming a saturated H$\alpha$ emission.} \\
\end{tabular}
%\tablefoottext{a}{$E^{\text{is}}(B-V) = E^{\text{tot}}(B-V) - 0.17$} \hfill
%\tablefoottext{b}{Assuming a saturated H$\alpha$ emission.}
\label{tab:extinct}
\end{table*}

\subsection{Spectral energy distribution}
\label{sect:sed}

In order to extend the characterisation of IGR J19294+1816, we
  will use the spectral energy distribution (SED) using the photometry
  on Table \ref{photometry}. It is well known, however, that free-free transitions
in the circumstellar envelope of the Be stars produce an excess
of photons at long wavelengths with respect to what can be expected
from the photospheric flux. 
The photometric points affected by this \emph{infrared excess} cannot be taken into account
in the SED fitting process.

\begin{figure}
  \centering
  \includegraphics[angle=-90,width=\columnwidth]{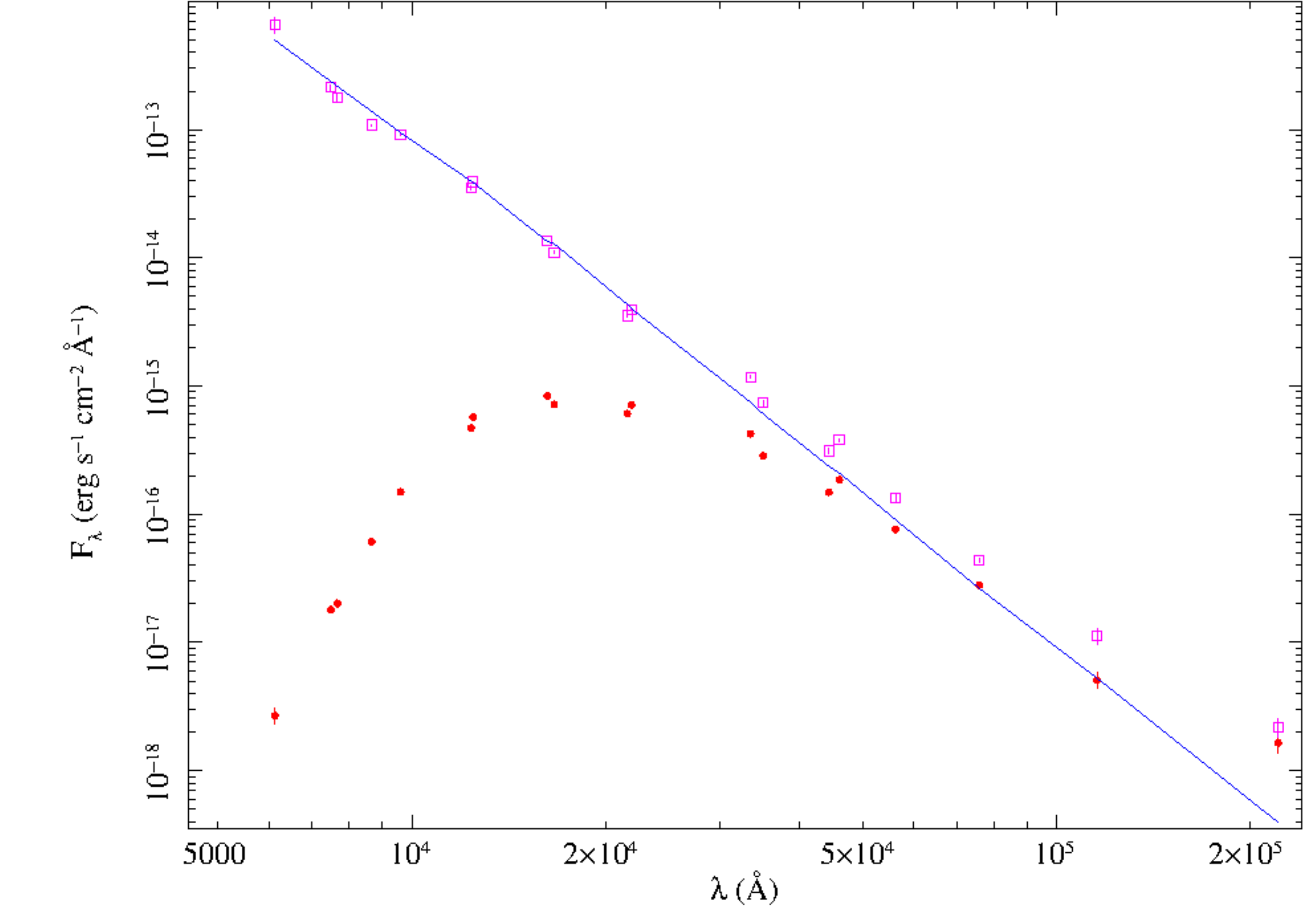}
  \caption{Spectral energy distribution of IGR J19294+1816. Filled circles represent
the observed flux, open squares refer to the de-reddened flux with
$E(B-V)=4.94$, and continuous line is the best fit model.
  }
  \label{fig:sed}
\end{figure}

The SED of the source is reported in Figure~\ref{fig:sed}. We built it using
the Virtual Observatory (VO) tool VOSA following the method described in
\citet{2008A&A...492..277B}\footnote{Photometric fluxes provided through VO services
\citep{2000A&AS..143...23O,2012ApJS..203...21A,2006AJ....131.1163S,
2000MmSAI..71.1123B,2010AJ....140.1868W}.}. %bayo et al. 2008
The filled (red) circles
represent the observed fluxes. The strong reddening is clearly
seen. At the wavelength of the $V$ filter (5500\,\AA), the SED flux
is of the order of $10^{-18}$ erg s$^{-1}$ cm$^{-2}$\,\AA$^{-1}$ which
corresponds to a magnitude of $V\sim 24$ mag, in agreement with the
estimation in the previous section. For de-reddening the SED, we use the
deduced $E(B-V)$ and the extinction law
by \citet{1999PASP..111...63F} %fitzpatrick 1999
improved by \citet{2005ApJ...619..931I} %indebetouw et al. 2005
in the infrared. The resulting de-reddened fluxes are shown as open
(magenta) squares. Finally, we fitted the SED with the TLUSTY
OSTAR2002+BSTAR2006 grid of stellar atmosphere models
\citep{1995ApJ...439..875H,2003ApJS..146..417L,2007ApJS..169...83L}. %tlusty references
%
%
%We note that, for BeXBs, the infrared photometry cannot be used
%at all to perform a SED fitting which aims at characterising the donor
%star because the circumstellar emission produces a strong IR excess.
%However, the fact that all three sources are consistent with
%BeXB systems allows us to constrain some key parameters in the model
%\citep{2012A&A...539A.114R}. %riquelme et al. 2012
%Moreover, VOSA take into account the error in magnitudes to calculate
%fluxes.
The
best fit (continuous, blue line) corresponds to a $T_{\mathrm{Eff}} =
30\,000$ K and $E(B-V)=4.94$. The
IR excess above 3 $\mu m$ is clearly seen (growing towards longer wavelengths). The
bolometric luminosity for the IR counterpart can now be estimated from the
ratio of observed to de-reddened fluxes\footnote{Changing the $log\, g$ had no
noticeable effect in our fits given the uncertainties and the lack of
data below 4000 \AA.}. This turns out to be $\approx 8\times 10^{37}$
erg s$^{-1}=(2.1\pm 0.5)\times 10^4\, L_{\sun}$. This is consistent with
a B0-2 star of luminosity class V.

\section{Discussion and conclusions}
\label{conclusion}

IGR J19294+1816 is a transient system with a likely B1Ve donor. The unabsorbed
X-ray flux (2--10 keV) reported by \citet{2009A&A...508..889R} %rodriguez et al. 2009
varies in the range of $(0.7-3.8)\times 10^{-11}$ erg\,s$^{-1}$\,cm$^{-2}$
following the activity of the source with \emph{INTEGRAL}, \emph{RXTE}
and \emph{Swift}, while it reaches $5.4\times 10^{-10}$ erg\,s$^{-1}$\,cm$^{-2}$
in outburst (20--50 keV energy band)
\citep{2011A&A...531A..65B}. %bozzo et al. 2011
At the estimated distance of 11 kpc, the X-ray luminosity would be
$(0.5-7.8)\times 10^{36}$ erg\,s$^{-1}$ in quiescence while the peak luminosity would be as high as $4.2\times 10^{37}$ erg\,s$^{-1}$,
assuming isotropic emission. The new NIR spectroscopy presented here
as well as the WISE photometry analysis strongly favour a
Be donor.  Hence, our results, combined with
the properties of the X-ray data, firmly classify IGR J19294+1816 
as a Be X-ray binary.

\section*{Acknowledgements}

  Part of this work was supported by the Spanish
  MINECO project number ESP2014-53672-C3-3-P. The authors
would like to thank the referee for her/his useful comments and suggestions.
  Based on observations made with the 
  Italian TNG operated on the island of
  La Palma by the Fundaci\'on Galileo Galilei of the INAF 
  (Istituto Nazionale di Astrofisica) at the Spanish ORM
  of the IAC.
  %This publication makes use of VOSA, developed under the Spanish Virtual
  %Observatory supported from the Spanish Ministry of Education
  %and Competitiveness (MINECO) through grant AYA2014-55216.
  This publication makes use of data products from the 2MASS,
  which is a joint project of the University of Massachusetts and
  the IPAC/CALTECH, funded by the NASA and the NSF.
  This publication makes use of data products from the WISE
  and NeoWISE, which are joint projects of the UCLA,
  and the JPL/CALTECH, funded by the NASA.
  This research has made use of the SIMBAD database, operated at CDS, Strasbourg, France.
  The UKIRT Wide Field Camera project is described in \citet{2007A&A...467..777C}.
  This work has made use of data from the European Space Agency (ESA)
  mission {\it Gaia} (\url{https://www.cosmos.esa.int/gaia}), processed by
  the {\it Gaia} Data Processing and Analysis Consortium (DPAC,
  \url{https://www.cosmos.esa.int/web/gaia/dpac/consortium}). Funding
  for the DPAC has been provided by national institutions, in particular
  the institutions participating in the {\it Gaia} Multilateral Agreement.
  J.J.R.R. acknowledges financial support from the Generalitat Valenciana
  and University of Alicante projects GV/2014/088 and GRE12-35, respectively.
  A.M. acknowledges the support by the grant by the Vicerectorat d'Investigaci\'o,
  Desenvolupament i Innovaci\'o de la Universitat d'Alacant under visiting
  researcher programme INV15-10.

%%%%%%%%%%%%%%%%%%%%%%%%%%%%%%%%%%%%%%%%%%%%%%%%%%

%%%%%%%%%%%%%%%%%%%% REFERENCES %%%%%%%%%%%%%%%%%%

% The best way to enter references is to use BibTeX:

\bibliographystyle{mnras}
\bibliography{jjrr_tng02} % if your bibtex file is called jjrr_tng02.bib

% Alternatively you could enter them by hand, like this:
% This method is tedious and prone to error if you have lots of references
%\begin{thebibliography}{99}
%\bibitem[\protect\citeauthoryear{Author}{2012}]{Author2012}
%Author A.~N., 2013, Journal of Improbable Astronomy, 1, 1
%\bibitem[\protect\citeauthoryear{Others}{2013}]{Others2013}
%Others S., 2012, Journal of Interesting Stuff, 17, 198
%\end{thebibliography}

%%%%%%%%%%%%%%%%%%%%%%%%%%%%%%%%%%%%%%%%%%%%%%%%%%

%%%%%%%%%%%%%%%%% APPENDICES %%%%%%%%%%%%%%%%%%%%%

%\appendix

%\section{Some extra material}

%If you want to present additional material which would interrupt the flow of the main paper,
%it can be placed in an Appendix which appears after the list of references.

%%%%%%%%%%%%%%%%%%%%%%%%%%%%%%%%%%%%%%%%%%%%%%%%%%

% Don't change these lines
\bsp	% typesetting comment
\label{lastpage}
\end{document}